# PiEEG kit - bioscience Lab in home for your Brain and Body

Ildar Rakhmatulin, PhD, PiEEG

| | |
|---|---|
| **Website** | https://pieeg.com/ |
| **GitHub** | https://github.com/pieeg-club/PiEEG_Kit |
| **YouTube presentation** | https://youtu.be/hFEF7NFFbZ8 |

**Abstract**
PiEEG kit is a multifunctional, compact, and mobile device that allows measure EEG, EMG, EOG, and EKG signals. The PiEEG Box incorporates the Raspberry Pi-based PiEEG shield, an EEG electrode cap, a display screen, additional sensors about body parameters and the environment, and other necessary peripherals, software, and an SDK course to learn signal processing into a single, portable unit. This integrated solution addresses the need for a compact, user-friendly, and accessible EEG measurement tool for researchers and hobbyists. The PiEEG Box builds upon the open-source foundation of the original PiEEG device, offering 8-channel EEG recording capabilities. By combining all required elements into one package, the PiEEG Box significantly reduces setup time and complexity, potentially broadening the application of EEG technology in various fields including neuroscience research, brain-computer interfaces, and educational settings.

## 1. Introduction

Electroencephalography (EEG) has long been a cornerstone of neuroscience research and brain-computer interface (BCI) development. However, traditional EEG systems often present challenges in terms of cost, portability, and ease of use, limiting their availability and widespread adoption. The advent of low-cost single-board computers such as the Raspberry Pi has opened up new opportunities for more accessible EEG solutions. Building on this potential, we previously introduced the PiEEG [1], a Raspberry Pi-based EEG data collection system. Although PiEEG significantly lowered the barrier to entry for EEG experiments and research, users still faced the challenge of assembling the various components to create a complete EEG measurement setup.

To address this limitation, we introduce PiEEG kit, a complete EEG measurement solution that combines the PiEEG device with all the necessary components for collecting brain signals. PiEEG kit represents a significant step forward in making EEG technology more accessible, user-friendly, and portable.

## 2. Main characteristics of PiEEG Box

**Integrated Design:** The PiEEG Box integrates the PiEEG Shield, Raspberry Pi, EEG cap, display screen, etc into one compact package.

**Portability:** Its versatility makes the PiEEG Box highly portable, facilitating EEG measurements in a variety of settings, from labs to field studies.

**User-friendly Interface:** The built-in display and intuitive controls simplify the EEG recording process, making it accessible to users with varying levels of technical expertise.

**Expandability**: While offering a complete out-of-the-box solution, the PiEEG Box maintains the open source spirit of the software and provides an SDK, allowing for customization and expansion of the system.

**Cost-effective:** Using readily available components and open source software, the PiEEG Box provides a low-cost alternative to commercial EEG systems.

This article will detail the design and implementation of the PiEEG Box, discuss its technical specifications, and explore potential applications in research areas. We aim to demonstrate how this integrated solution can accelerate research

## 2. Review of related works

Previously, we created devices for recording biodata data using Raspberry Pi[1], Arduino [2] and Jetson Nano[3]. Also, in addition to a large number of innovative devices that are widely used in the development, low-cost solutions such as OpenBCI [4], solutions Arduino Mega2560[5], and many others are also known, a review of which is presented in the following works. But for the most part, in our opinion, these devices require certain programming and technical skills to set up biodata reading. In this solution, a case is presented that allows you to quickly begin studying bio data.

## 3. Technical Realizations

The basis of this device was taken device PEEG and improved format all in one. Therefore in this case it is necessary only to open the box connect (not included components) the power supply (5V battery), keyboard, and mouse, and from this moment launch the software and start reading EEG data. A general view of the device is presented in Figure 1.

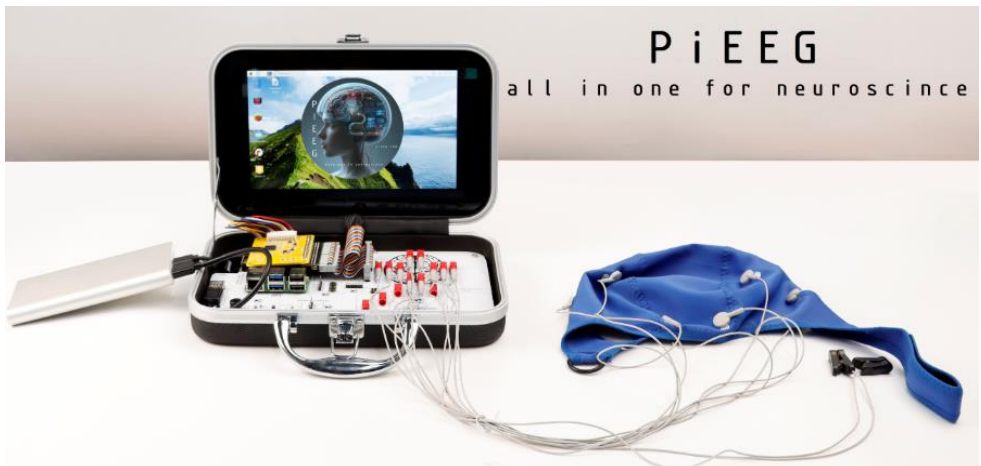

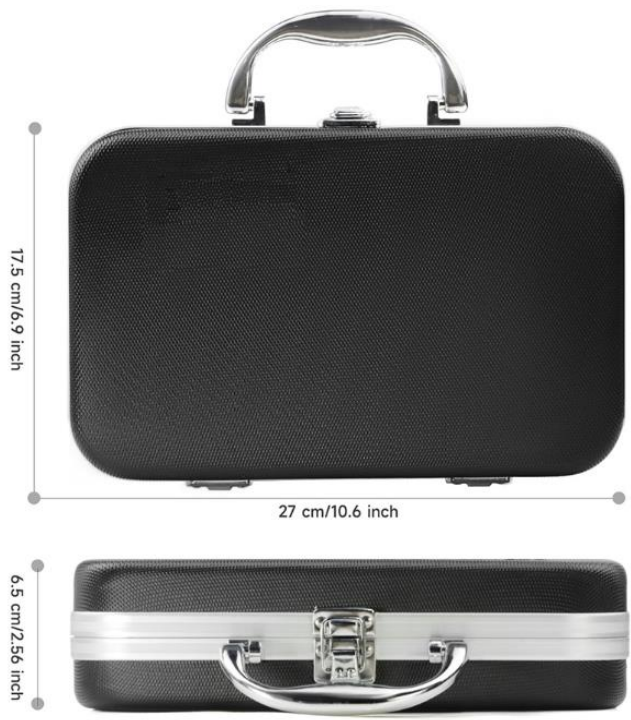

Fig. 1. General view of the PiEEG kit

The board has advanced functionality - this is the reading of EEG, EMC, ECG and additional sensors, Figure 2 shows the functionality of the board.

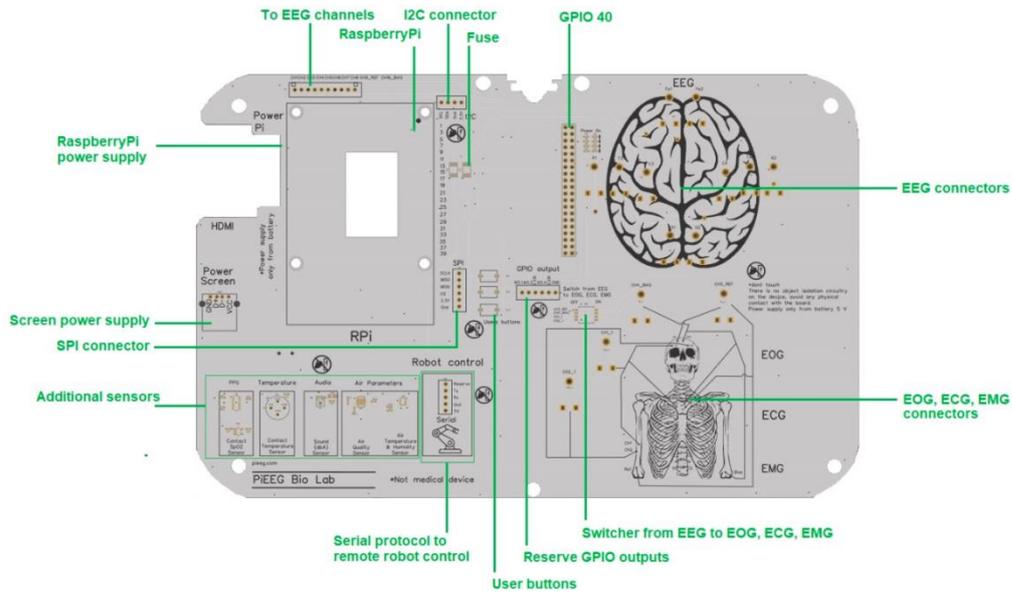

Fig 2. Functionality of the PiEEG kit

We integrated also additional sensor to detect paraments of human body and env to increase opportunity of feature extraction for EEG, Figure 3.

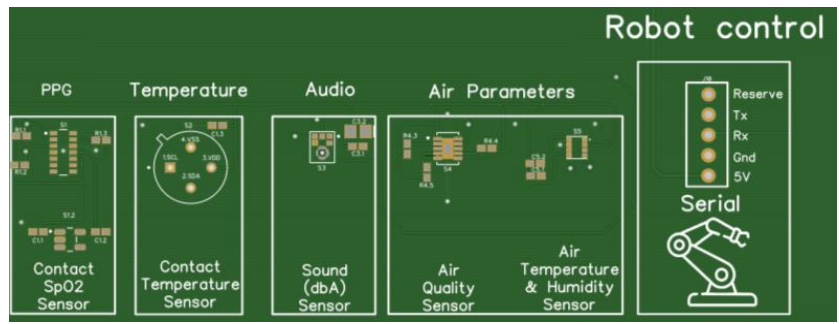

Fig.3. Sensors board and connector for Serial transmitter to Robot control

The heart of this device is the PiEEG shield, which uses the ADS1299 analog-to-digital converter from Texas Instruments. The shield is connected to Raspberry Pi via GPIO 40. Figure 4 shows the PEEG device, which is based on a 24-bit ADC.

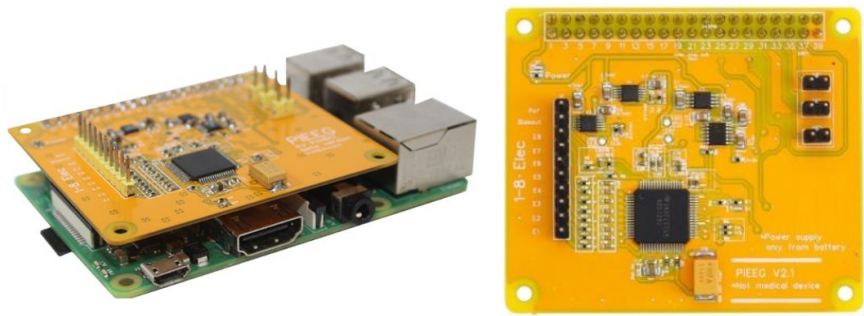

Fig. 4. General view of the PiEEG board

Figure 5 shows the schematic connection of the electrodes to the PIEEG and the meaning of the diodes on the board.

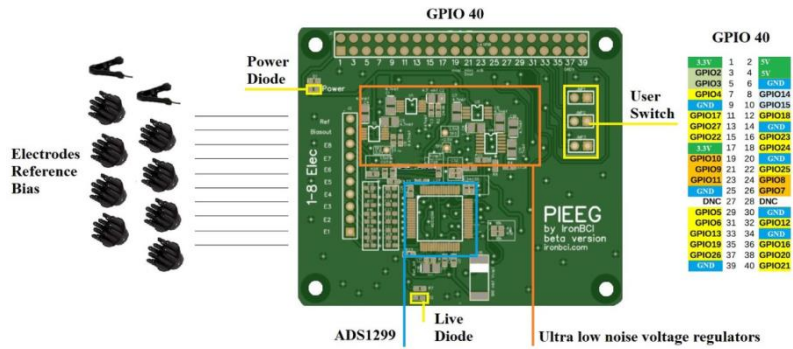

Fig. 5. Electrode connection diagram for PiEEG

## 4. Measurement test

We have made many EEG recordings with dry electrodes Ag/AgCl electrodes and tests, examples of recordings are available on GitHub (250 samples per second). The following graphs show recordings from the dry electrode from the Fz position in accordance with the International Electrode Placement System "10-20".

### 4.1 Electroencephalogram (EEG)

An electroencephalogram (EEG) is a non-invasive medical test that records the electrical activity of the brain. It provides valuable information about brain function and is used to diagnose a variety of neurological disorders. A reliable method of confirming that an EEG recording is occurring correctly is the alpha test. Figure 6 shows a recording with a sequence of 5 seconds of eyes closed, and 5 seconds of eyes open. Alpha waves (α waves), a type of brain signal that fluctuates between 8 and 12 Hz. These waves are typically found in awake people who close their eyes in a relaxed state.

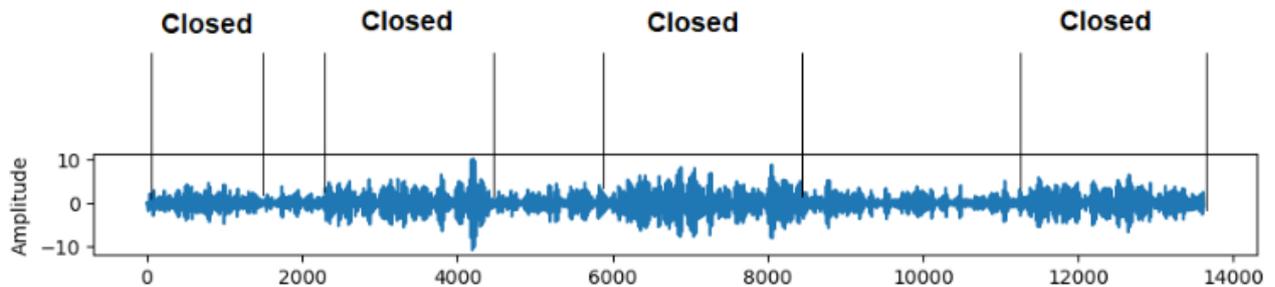

Fig. 6. Alpha rhythm test. Sequence. 5 sec eyes closed, 5 sec open. Electrode in position Fz. Dry electrodes Ag/AgCl at 250 samples per second.

### 4.2 Electrooculography (EOG)

Electrooculography is a diagnostic technique used to measure the electrical potential difference between the front and back of the human eye, known as the corneo-retinal constant potential. This technique produces a signal called an electrooculogram (EOG), which is primarily used in ophthalmology to evaluate eye movements and diagnose conditions related to the retinal pigment epithelium. Figure 7 shows the recording process of chewing signals (4-3-2-1 times) and EOG signals during blinking (4-3-2 times).

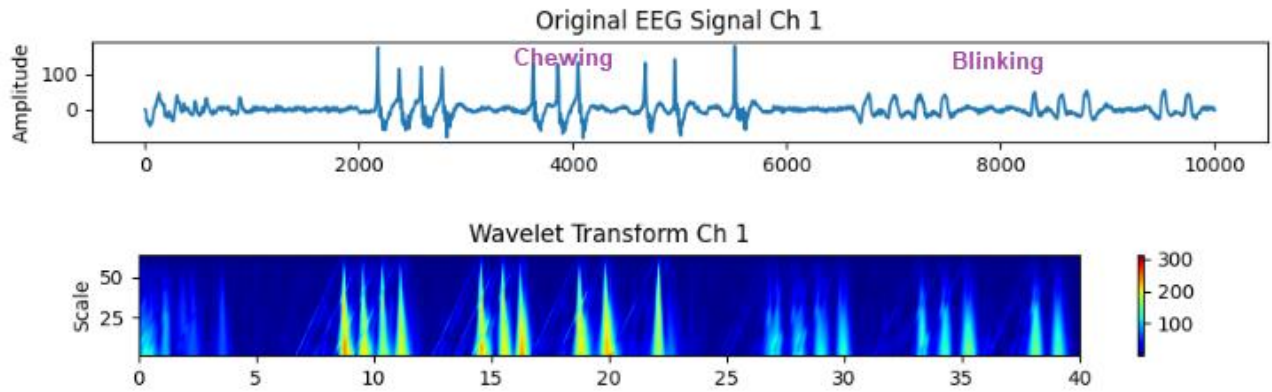

Fig.7. Example of blinking and chewing artifacts. Chewing 4-3-2-1 times and then blinking also 4-3-2 times. Electrode in position Fz. Dry electrodes Ag/AgCl at 250 samples per second.

### 4.3 Electromyography (EMG)

Electromyography (EMG) is a diagnostic procedure that evaluates the health and function of muscles and the nerve cells (motor neurons) that control them. It measures the electrical activity produced by skeletal muscles, providing valuable information about muscle and nerve function. Figure 8 shows the dataset for taking the EMG signal from the right arm biceps when clenching and opening the palm into a fist.

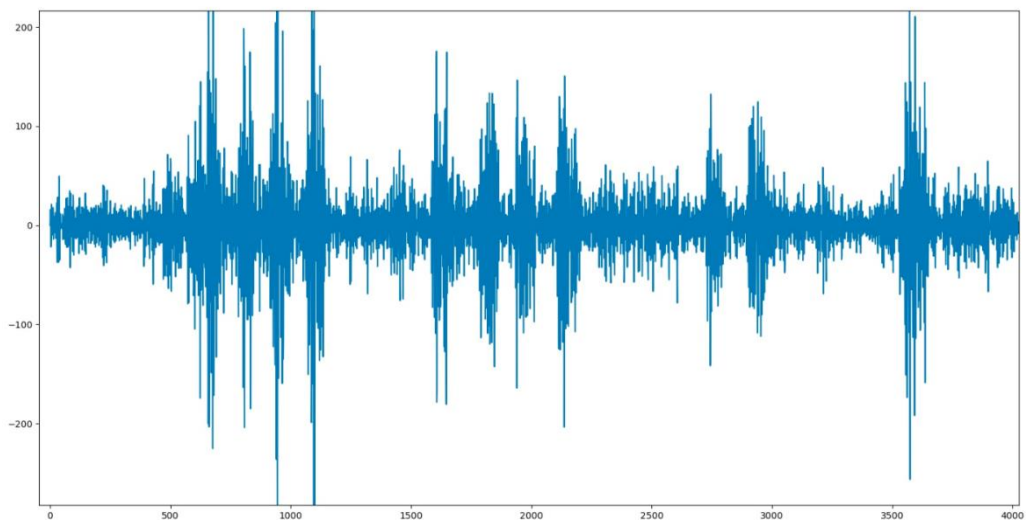

Fig. 8. Example of EMG signal recording with electrodes Ag/AgCl

### 4.4 Electrocardiogram (ECG)

An electrocardiogram (ECG) is a simple test that records the electrical activity of the heart. It provides valuable information about how well the heart is working. Figure 9 shows the part of the dataset taken while the ECG recording.

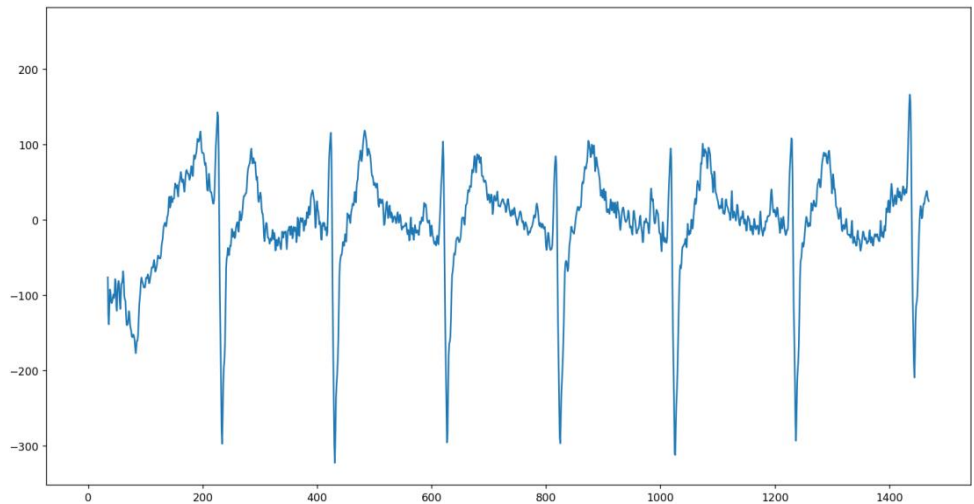

Fig. 9. Example of ECG signal recording with electrodes Ag/AgCl

## 4. Software
### 4.1. Graphic unit interface GUI
The PiEEG kit has software installed for data visualization. The software allows change of data reading parameters, and most importantly, full access to the ADS1299 registers, which makes it possible to change the gain of the EEG signal due to the internal ADS1299 amplifiers, the speed of data transfer, and much more. A screenshot of the software is shown in Figure 7.

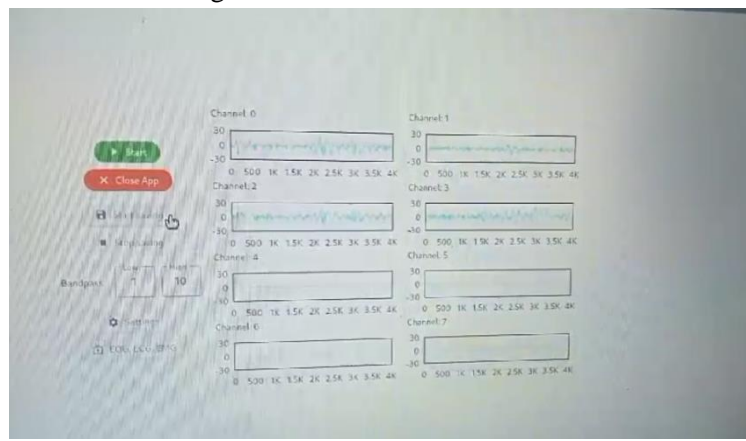

Fig.7 Screenshot of software for working with real-time data

### 4.2 Software development kit (SDK)
An open-source SDK for researchers, allowing them to create their own scenarios for data analysis and feature extraction from EEG signals. The SDK is located on GitHub at https://github.com/pieeg-club/PiEEG_Kit/tree/main/SDK. We provide Python scripts for data visualization and data saving.

### 4.3 Course included to PiEEG
We offer a comprehensive course on signal processing for EEG data, demonstrating the simplicity and effectiveness of processing EEG data using Python. The course also includes practical examples of applying EEG data analysis to real-world tasks, Figure 8.

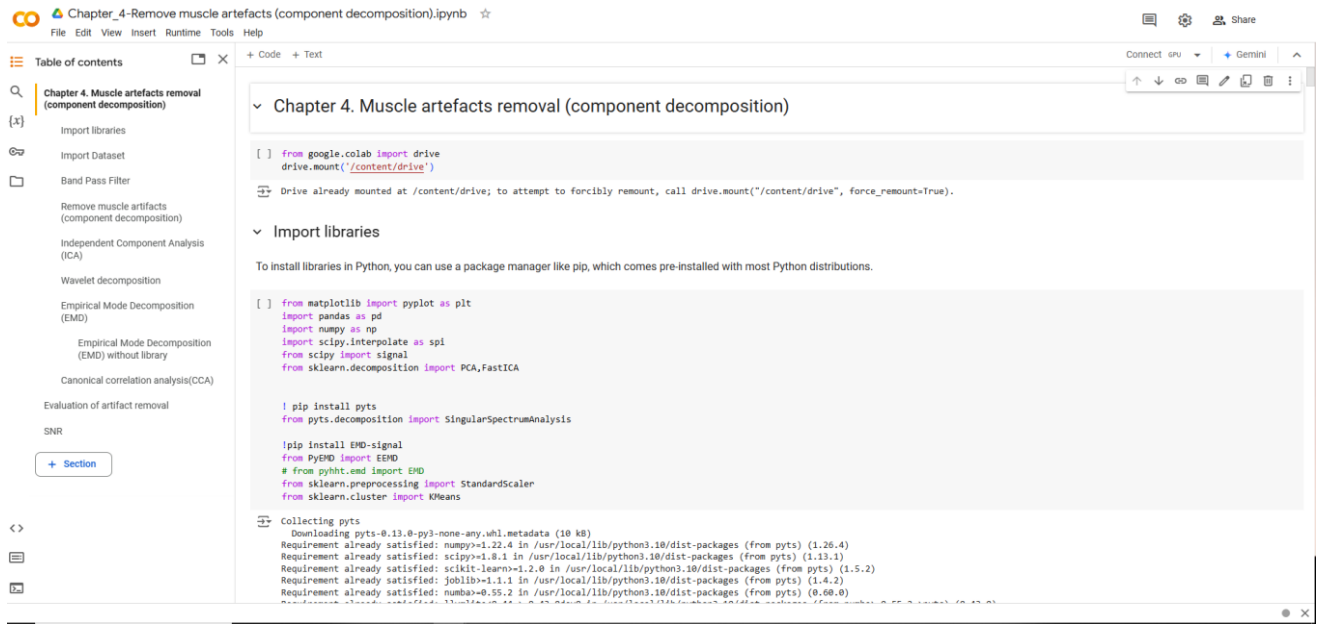

Fig.8. Screenshot of Chapter 4 of the Artefact Removal Course

**Conclusion and Prospect**

The device is intended for studying neuroscience. But the potential application of the PEEG kit device goes far beyond the scope of our initial idea, Figure 9.

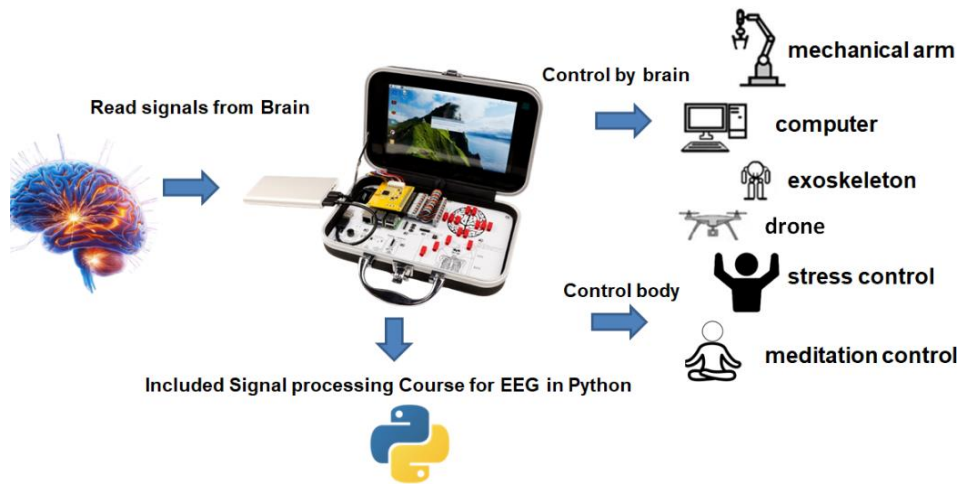

Fig.9. Potential application of the PiEEG kit

**Democratizing Neuroscience: Our Mission and Journey**

At the heart of our endeavor lies a singular vision: make neuroscience accessible to everyone. Our journey began with the development of the PEEG device, a pioneering step towards simplifying bio-data measurement. However, we soon recognized that even this solution presented challenges for those without a background in electronics, particularly in connecting the Raspberry Pi to the PEEG.

**Evolving Our Approach**

In response to these challenges, we've created an all-in-one solution that streamlines the process and eliminates connectivity issues. But our commitment extends beyond merely providing tools for data collection. We've developed a comprehensive, instructive course that bridges the gap between theory and practice, demonstrating how neuroscience principles can be applied in everyday life.

**Looking to the Future**

Our passion for advancing neuroscience drives us to seek collaborations and push the boundaries of what's possible. We're particularly excited about developing algorithms to extend artifact detection and removal, including motor artifacts. These innovations have far-reaching potential, especially in the area of healthcare, virtual reality and EEG integration.

**Open Invitation**

By making neuroscience more approachable and applicable, we're not just advancing a field of study – we're empowering individuals to understand and harness the power of their own minds.

**Important**

It's important to note that PiEEG is not classified as a medical device. For a comprehensive understanding of its intended use and limitations, it should review our detailed liability information

https://pieeg.com/liability-pieeg/